%% file: main.tex
\documentclass[a4paper]{article}
\usepackage[margin=1in]{geometry} % full-width
\usepackage{amsmath}
\usepackage{amsthm}
\usepackage{amssymb}

\usepackage{booktabs}

\usepackage[utf8]{inputenc}
\usepackage{hyperref}
\hypersetup{
	unicode,
	pdfauthor={Author One, Author Two, Author Three},
	pdftitle={A simple article template},
	pdfsubject={A simple article template},
	pdfkeywords={article, template, simple},
	pdfproducer={LaTeX},
	pdfcreator={pdflatex}
}

\usepackage[sort&compress,numbers,square,super]{natbib}
\usepackage{graphicx}
\usepackage[labelfont=bf, font=footnotesize]{caption}
\usepackage{indentfirst}

\usepackage{listings}
\usepackage{xcolor}
\definecolor{codegreen}{rgb}{0,0.6,0}
\definecolor{codedarkblue}{rgb}{0.039,0.188,0.411}
\definecolor{codegray}{rgb}{0.5,0.5,0.5}
\definecolor{codepurple}{rgb}{0.58,0,0.82}
\definecolor{backcolour}{rgb}{0.95,0.95,0.92}
\definecolor{backcolourGray}{rgb}{0.97,0.97,0.97}

\lstdefinelanguage{json}
{
    morestring=[b]",
    morestring=[d]',
    string=[s]{"}{"},
    comment=[l]{:\ "},
    morecomment=[l]{:"},
}

\lstdefinestyle{mystyle}{
    backgroundcolor=\color{backcolourGray},   
    commentstyle=\color{codegreen},
    keywordstyle=\color{magenta},
    numberstyle=\tiny\color{codegray},
    stringstyle=\color{codedarkblue},
    basicstyle=\footnotesize\ttfamily,
    breakatwhitespace=false,         
    breaklines=true,                 
    captionpos=b,                    
    keepspaces=true,
    numbers=left,                    
    numbersep=2pt,                  
    showspaces=false,                
    showstringspaces=false,
    showtabs=false,                  
    tabsize=2
}
\usepackage{comment}

\title{\textbf{CateCom: a practical data-centric approach to \\ categorization of computational models.}}
\author{Alexander Zech$^{1,2}$ and Timur Bazhirov$^1$}

\date{
	$^1$Exabyte Inc., San Francisco, CA, United States \\ 
	$^2$University of California, Berkeley, Berkeley, CA, United States \\[1.2ex]
	\today \\
	\texttt{timur@exabyte.io} \\
}

\begin{document}
	\maketitle

% --- A B S T R A C T ---	
\begin{abstract}
	The advent of data-driven science in the 21st century brought about the need for well-organized structured data and associated infrastructure able to facilitate the applications of Artificial Intelligence and Machine Learning. We present an effort aimed at organizing the diverse landscape of physics-based and data-driven computational models in order to facilitate the storage of associated information as structured data. We apply object-oriented design concepts and outline the foundations of an open-source collaborative framework that is: (1) capable of uniquely describing the approaches in structured data, (2) flexible enough to cover the majority of widely used models, and (3) utilizes collective intelligence through community contributions. We present example database schemas and corresponding data structures and explain how these are deployed in software at the time of this writing.
		
	\textbf{Keywords:} data structures, data standards, artificial intelligence, machine learning.
\end{abstract}

%\tableofcontents

\input{01-introduction/introduction}

\input{02-methodology/methodology}

\input{03-examples/examples}

\input{04-discussion/discussion}

\input{05-conclusion/conclusion}

% --- B I B L I O G R A P H Y ---
\bibliography{refs}
	
% --- A P P E N D I X ---
\appendix
%\renewcommand{\thesection}{\Alph{section}}
%\numberwithin{equation}{section}
\clearpage
\renewcommand\thesection{\Alph{section}}
\renewcommand\thelstlisting{\thesection.\arabic{lstlisting}}
\setcounter{lstlisting}{0}
{\noindent \Large\textbf{Appendix}}
\section{Schema Examples}
\label{app:1}

\input{99-supporting-info/supporting_information}

\end{document}

%% file: 01-introduction/introduction.tex
\section{Introduction}
\label{sec:intro}
    The proliferation of data-driven science creates the need for systematically organized and machine-readable data formats.\cite{Baker2016-mt} While in the past considerable effort has been dedicated to structuring simulation results, the organization of simulation metadata has only recently gained attention. One major aspect of such an undertaking is the classification of numerous computational models. Early forms\cite{Pople1965-yo, Karplus1990-aj, Tarczay2001-zm} of the categorization of quantum chemical models were based on only a few distinguishing descriptors, i.e. the treatment of electron correlation and one-particle basis set, as well as the type of Hamiltonian. More recently, projects emerged which not only collect results and metadata from output files of simulation packages but also define database schemas for their storage. For instance, the Novel Materials Discovery (NOMAD)\cite{Draxl2018-mm} repository includes a structured collection of computational model metadata as part of its \texttt{metainfo} component. Another example is QCSchema\cite{molssi} by the Molecular Science Software Institute (MolSSI), which provides software-independent data structures for quantum chemistry geared towards unified and consistent workflows. Organizing the computational models and their results can also be achieved through an ontology, often expressed in the Web Ontology Language (OWL)\cite{owl2}. One such example is the OntoCompChem ontology\cite{Krdzavac2019-ji}, which is applied to quantum chemistry calculations as part of the MolHub\cite{Phadungsukanan2012-ze} web service. In the domain of materials science, ontologies are more prevalent but often focus on specific subdomains such as nanoparticles\cite{Thomas2011-fz}. There are, however, examples of general ontologies, such as the Elementary Multiperspective Material Ontology (EMMO)\cite{EMMO} or the Materials Design Ontology (MDO)\cite{Li2020-rf}. There exist a number of NIST/MGI-led prior efforts where large-scale high-throughput computational approaches have been used to screen thousands of compounds with subsequent web-based dissemination, databasing, and data-mining. \cite{Curtarolo2012-ik, Jain2013-va, Saal2013-ko, Calderon2015-vr, Kirklin2015-qz, Draxl2018-mm, Choudhary2020-uc, Huber2020-mv}
    
    Most current data structures for computational models include little information beyond just its name relying on the description in the scientific literature. Such an approach makes it difficult to construct data-driven predictions. We elaborate on the existing approaches by constructing a framework able to utilize previously obtained data (also allowing the generation of new data) to categorize the important descriptive features for a set of entities (materials, simulation workflows/models, computational methods) and target properties of interest (electronic, chemical, thermodynamic, structural properties) to construct associative maps, and organize "actionable" data in this extremely diverse and complex domain.
    
    Our effort follows an object-oriented approach by building a basis of unit models, which are small inseparable sets of equations pertaining to a specific physical description of reality (e.g. Kohn-Sham Density Functional Theory). A given level of theory may further be expressed as a combination of unit models. Such modularity allows us to cover a diverse range of use cases. In addition to the categorization of the computational models, we discuss a semantic layer in the form of an ontology, which not only facilitates a more accurate description of the relationships between models, but also provides the foundation for further applications (e.g. building a knowledge graph, improved search, or AI/ML engines). The design of the proposed data structures is also coupled with the application thereof inside an online software platform\cite{exabyte} allowing to create a very short feedback loop and improve the resulting implementation based on feedback from thousands of users of the platform. Our standards facilitate the development of artificial intelligence tools that can reduce the dimensionality and complexity of the research work in materials science and chemistry, with the aim of eventually enabling inverse design. Our goal is to build collective intelligence utilizing contributions from a large audience of materials scientists (well beyond the select few experts in the computational field) and chemists in a controllable and high-level fashion.
    
    The following section of this paper will introduce the simulation entities and the principles on which the model categorization scheme is based. The subsequent section illustrates the representation of the entities as data structures using a set of selected examples. In the fourth section, we discuss the relevancy and limitations of the categorization scheme and propose a community-driven approach to extend the categorization scheme. Finally, the principal conclusions are presented together with a perspective on future applications.

%% file: 02-methodology/methodology.tex
\section{Methodology}
\label{sec:methodology}

%%% The concept of CateCom is here
% - wishlist of classification properties (maybe write summary instead)
% - where is it implemented? -> ESSE, or perhaps move to new repo? (better if someone is just interested in ESSE ontology)
% - insert diagram
% - technical details (JSON schema, etc.)
% - list classification rules
% - model graph?
% - method (connect to precision & accuracy)
%
% data-centric categorization scheme
%A categorization scheme should comprise the following properties in order to accommodate the variety of computational models:
%(x) object-oriented ansatz for reusability,
%(x) 
%(x) set of rules in order to cover a broad range of cases,
%(x) flexibility towards modifications of a theory that do not warrant a separate object.

    \subsection{General Approach}
    \label{subsec:methodology-general}
    
\begin{figure}[!ht]
            \centering
            % https://drive.google.com/file/d/1pn-fv0flaeVRdFdOiQRVYaqjhg2argRR/view?usp=sharing
            \includegraphics[width=0.6\textwidth]{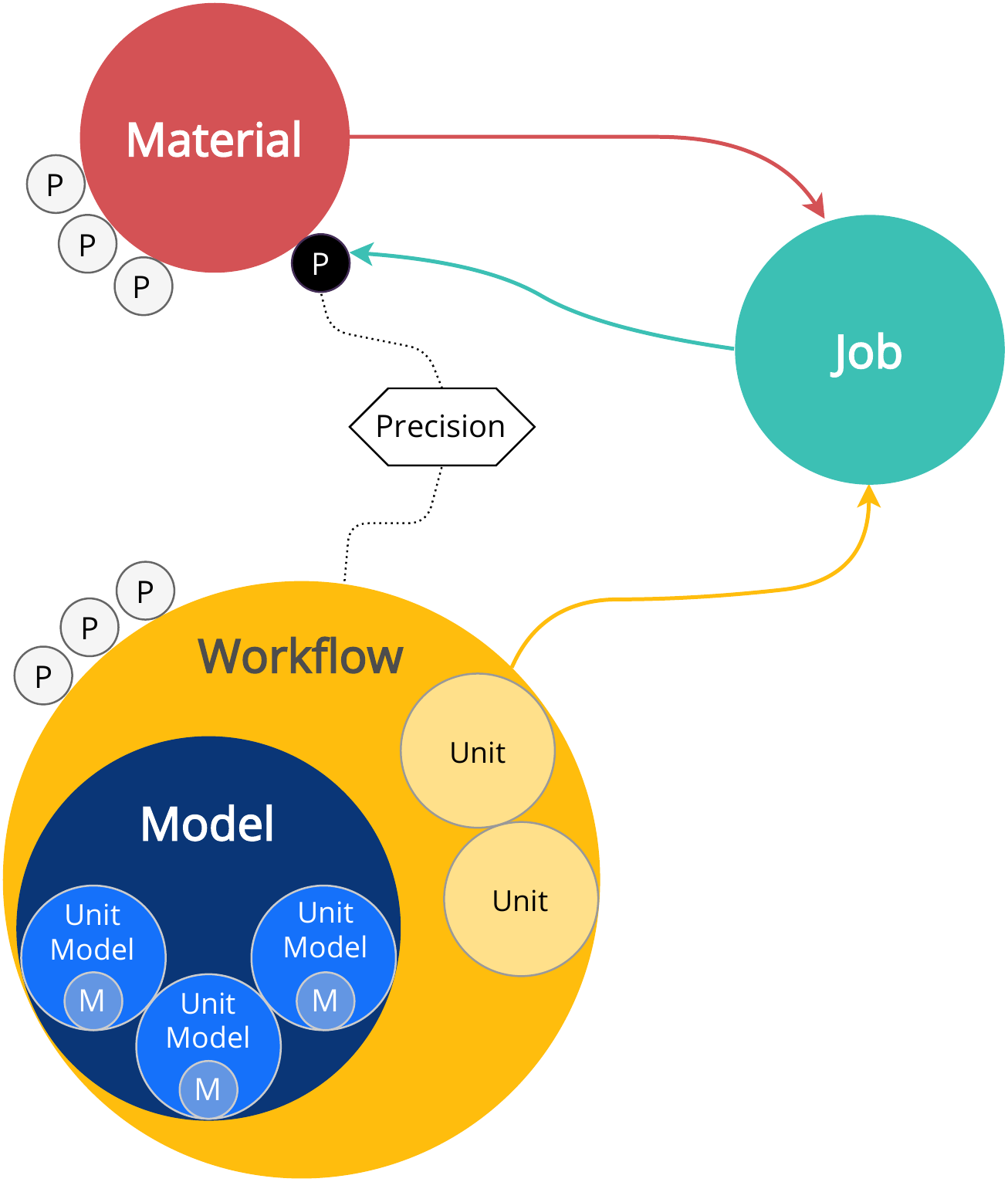}
            \caption{
                A visual representation of the key Entities considered in the current approach and relationships between them. A Job represents a research (computational) task involving a combination of Workflow and Material(s) Entities in order to produce initial \textit{a priori} (gray) and target \textit{a posteriori} properties (black) denoted by the letter "P". Resulting properties have a certain Precision. A Workflow is further made of individual workflow Units responsible for elementary computational operations (e.g. a "one-shot" run of simulation software). A Workflow has a (compound, in the general case) Model associated with it, which in turn is composed of Unit Models. The categorization of such Unit Models represents the focus of the current work.
            }
            \label{fig:methodology:entities}
        \end{figure}

        On an abstract level, we choose to represent a computational simulation in the form of several key entities (see Fig.~\ref{fig:methodology:entities}). First of all, the \textit{material} entity defines the chemical composition of the system under investigation. Although this entity is termed \textit{material}, it may also constitute a non-periodic molecular species. The \textit{workflow} entity organizes the sequence of tasks, for instance, execution of simulation software or input/output operations. The workflow also includes the specification of the theoretical model, which in turn is represented by the \textit{model} entity. It should be noted that both the workflow and model entities are composed of reusable units, which may be combined in various ways. Applying the workflow on a material in order to produce one or more properties, i.e. connecting the workflow and materials entities, is achieved by the \textit{job} entity. This entity does not only serve as a container for material and workflow entities but also stores more technical information related to high-performance computing and resource allocation. Properties may either be derived from existing entities (gray) or occur as a direct result of the job (black). In the latter case, one can associate a \textit{precision} entity that is derived from the workflow. An important factor in determining the precision is the practical approach for solving the theoretical model, which is represented by the \textit{method} entity (M). As part of a model (or unit model), it stores the selection of algorithms, thresholds, and other practical parameters. 
        For the utilization on the Exabyte.io platform\cite{exabyte} and document-based, NoSQL (not only structured query language) databases in general, these entities are formulated as database schemas. Document-based, NoSQL databases are a convenient choice for such an application due to a few advantageous properties: (a) data that is accessed together is stored together (as opposed to joined from multiple data tables), (b) the organization of data can be as complex as one chooses thus supporting parent-child hierarchical structures, (c) data structures are not fixed and may be changed in response to new data models. The proposed database schemas have been implemented in the Exabyte Source of Schemas and Examples (ESSE)\cite{ESSEgithub} using the JSON Schema notation (Draft-04)\cite{JSONSchema}.

    \subsection{Components and Entities}
    \label{subsec:methodology-entities}

        The \texttt{ESSE} module comprises several main schemas for simulation data, such as workflow, material, property, method, and model. The following sections focus on the latter two entities, while all other schemas are briefly summarized in Sec.~\ref{subsubsec:methodology-entities-other}.
        
        \subsubsection{Model}
        \label{subsubsec:methodology-entities-model}
\begin{figure}[!ht]
                \centering
                \includegraphics[width=0.95\textwidth]{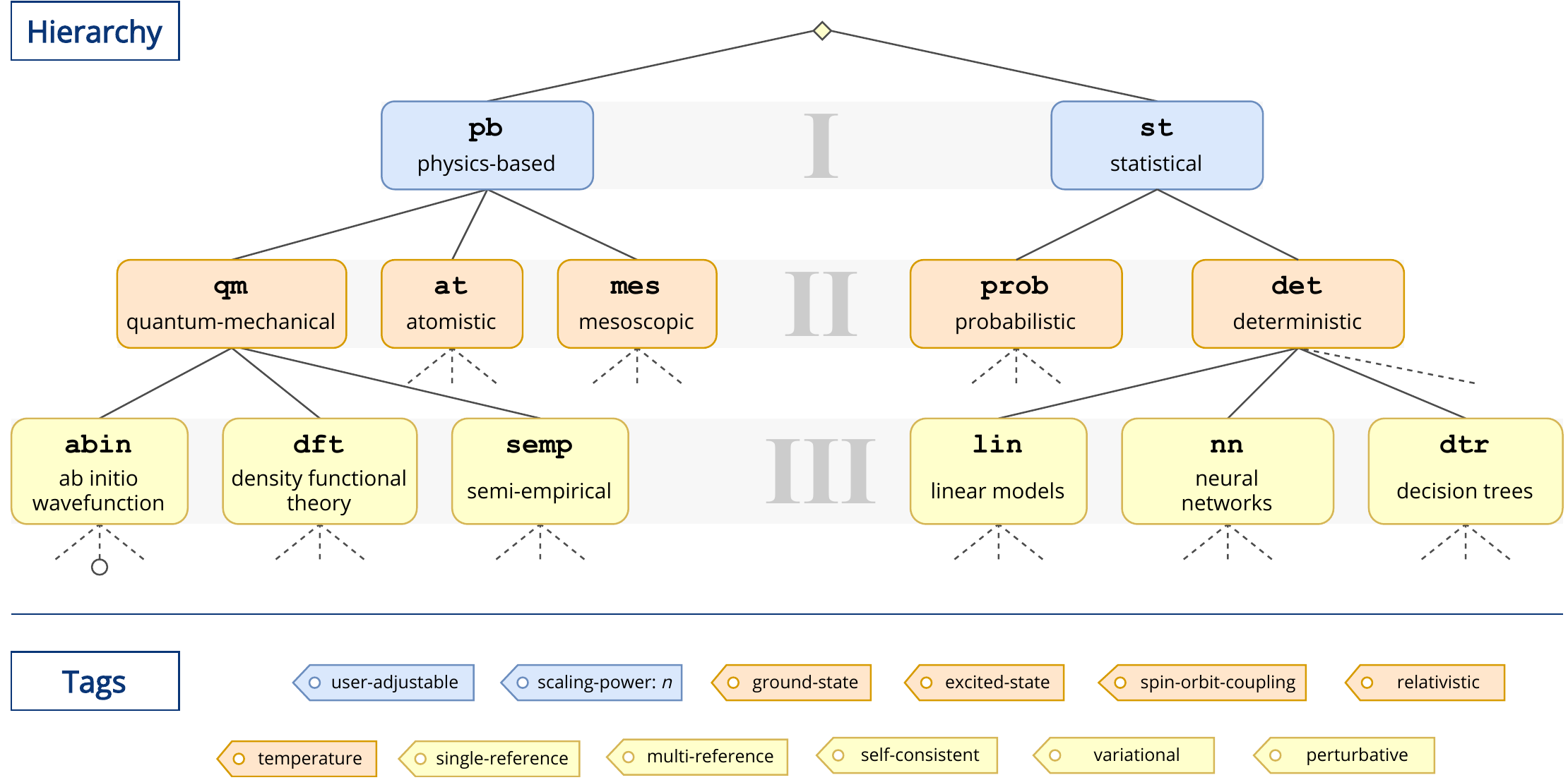}
                \caption{Schematic representation of the tiers of the CateCom categorization. Tier I to tier III of the unit model classification hierarchy and descriptive tags colored according to the tier in which they first appear. }
                \label{fig:model_tiers}
            \end{figure}

            The function of the model schema is to define a given computational model as accurately as possible and to simultaneously store all of the necessary metadata.
            % introduction of unit models & classification tree
            We chose to represent a given computational model in terms of one or more reusable, independent components, which we will refer to as \textit{unit models} in the further course. The final model which is applied to a system is a combination of said unit models and termed \textit{compound model}.
            % unit model definition
            We define a unit model as the smallest, logically consistent set of equations or operators associated with a central property (e.g. electronic energy).
            % discuss unit model overlap
            In practice, a unit model may not always be unambiguously defined, i.e such a case may require further partitioning into a set of unit models that would go beyond the scope of this classification scheme. %One therefore needs to balance exactness against pragmatism.
            The classification, therefore, requires a subtle balance of exactness and pragmatism.
            % Classification tree
            The classification tree involves three main tiers (Figure~\ref{fig:model_tiers}) in order to presort the models into families of models, for instance \textit{quantum mechanical} or \textit{classical}.
            Following tier III the models are further divided into more specific categories, which are also organized hierarchically using \texttt{type}, \texttt{subtype} specifiers.
            The design of the schemas follows an object-oriented approach whereby schemas share fields through inheritance (by means of the \texttt{allOf} keyword).
            At each level of the classification tree, a given schema thus includes all categorization specifiers of the preceding levels and may serve as a prototype for the following level.
            One advantage of this concept is that changes to categories or implementation of new categories only occur locally and are propagated automatically to the lower levels.
        
            % further structure of unit models
            %In addition to the classification specifiers, each unit model comprises \texttt{tag}, \texttt{modifier} and \texttt{augmentation} fields, which are also passed on through the categorization hierarchy (see Table~\ref{tab:methodology:labels} for examples).
            In addition to the classification specifiers, each unit model comprises a \texttt{tag} field, which is also passed on through the categorization hierarchy (see Table~\ref{tab:methodology:labels} for examples). The tags describe attributes of the unit model not included in the categorization and indicate whether a modifier or augmentation has been applied to the unit model.
            We define a modifier as an addition to a model, which expands upon the underlying physical principle without fundamentally changing the working equations (e.g. in a linear fashion). An example of a modifier is the inclusion of an additional external potential (e.g. due to point charges). An augmentation, on the other hand, is defined as an addition to a model, which does not change the underlying physical principles of the model. For instance augmentations include acceleration techniques such as resolution-of-the-identity or localization schemes (e.g. Edmiston-Ruedenberg localization\cite{Edmiston1963-pg}). Apart from the specifications above, the \texttt{tags} field may also hold user-defined labels.
            %Labels that do not fit in the two preceding specifications are collected in the tags container, which may also hold user-defined labels.
            %Tags do not relate to the model per se, but may hold user-defined labels.
            
            % Classification rules
            \begin{itemize}
                \item[Tier I$\;$ $\;$] At the primary level unit models are distinguished between \textit{physics-based} (\texttt{pb}) and \textit{statistical} (\texttt{st}).
                The latter category pertains to data-driven approaches which employ statistical relations in order to predict a result.
                Models based on fundamental laws of physics are assigned to the \textit{physics-based} category even if the unit model heavily relies on statistical elements (see also Sec.~\ref{subsec:methodology-classification}).
                
                \item[Tier II$\;$] Within the \textit{physics-based} group, the quantum mechanical (\texttt{qm}), atomistic (\texttt{at}) or mesoscopic (\texttt{mes}) category are used depending on the type of particle represented in the equations of the unit model. For instance, Kohn-Sham density functional theory (KS-DFT) naturally falls into the \texttt{qm} category due to its explicit dependence on electronic coordinates, while a force field such as CHARM22\cite{MacKerell1998-hb} only depends on atomic variables and is thus assigned to \texttt{at}. In the \textit{statistical} group a unit model falls into the probabilistic (\texttt{prob}) category if the predicted result incorporates some aspect of random variation, whereas the deterministic category (\texttt{det}) is chosen if it does not.
                % Example!
                
                \item[Tier III] % qm group -> abin, dft, semp
                The quantum-mechanical models are further divided into three categories. The \textit{ab initio} category (\texttt{abin}) comprises first-principle wavefunction models, such as Hartree-Fock theory, many-body perturbation theory, or coupled cluster theory, which do not require additional information about the system. With the electron density as the central quantum mechanical descriptor, the realizations of density functional theory are collected in the \texttt{dft} category. The \textit{semi-empirical} category (\texttt{semp}) contains parametrized quantum-mechanical models, which usually only describe valence electrons for computational efficiency.
                % statistical model branch
                As for the statistical model branch, Figure~\ref{fig:model_tiers} shows how the deterministic models can be further subdivided using the example of three prominent machine learning model categories. Linear models (\texttt{lin}) span the space of models which assume a linear relationship between the input variables ($\mathbf{X}$) and the dependent variable ($\mathbf{y}$). The neural network category (\texttt{nn}) contains all models that are based on a neural network architecture, i.e. a network of interconnected processing nodes whereby connections between nodes are represented by weights. The decision tree category (\texttt{dtr}), on the other hand, comprises models which can generate a prediction based on recursively splitting the dataset into subsets. The outcome of such a procedure is then a linear acyclic graph of decision nodes and 'leaves' (endpoints, which do not split the data any further). The decision tree approach is usually applied in an ensemble (random forest model), which in the CateCom scheme is represented as a compound model. Other examples of deterministic models not explicitly shown in Figure~\ref{fig:model_tiers} support vector machines or clustering algorithms, such as k-means.
                \end{itemize}

            % det group
            
            % list of tags, augmentations, modifiers
            \begin{table}[!ht]
                \centering
                \caption{Selected model attributes and the corresponding categorization levels.}
                \label{tab:methodology:labels}
                \begin{tabular}{l|p{9.5cm}|l}\toprule
                    Label & Explanation & Category \\\midrule
                    relativistic        & Inclusion of relativistic effects. & \texttt{pb}\\
                    user-adjustable     & The model contains additional parameters to fine-tune results. & \texttt{pb}\\
                    scaling-power:$n$  & The model exhibits a formal scaling of $n$-th power. & \texttt{pb}\\ 
                    self-consistent     & Non-linearity in the model is solved through self-consistent optimization. & \texttt{pb/qm}\\
                    temperature         & The model describes non-zero temperature effects. & \texttt{pb/qm}\\
                    excited-states      & Access to electronically excited states. & \texttt{pb/qm}\\
                    %ground state        & \\
                    spin-orbit coupling & The model accounts for spin-orbit coupling. & \texttt{pb/qm}\\
                    variational         & The model follows the variational principle. & \texttt{pb/qm}\\
                    single-reference    & The wavefunction is based on a single reference determinant. & \texttt{pb/qm}\\
                    multi-reference     & The wavefunction is based on multiple reference determinants. & \texttt{pb/qm}\\
                    perturbative        & The model contains elements of perturbation theory. & \texttt{pb/qm/abin}\\
                    \bottomrule
                \end{tabular}
            \end{table}

        \subsubsection{Method}
        \label{subsubsec:methodology-entities-method}
            While the unit model pertains to the accuracy of a computational simulation, the \textit{method} concerns its precision. The CateCom collection, therefore, includes a \texttt{method} schema for parameters concerning the computational methodology, such as convergence thresholds or hyperparameters (machine learning). As it is closely related to a model, method schemas are part of unit models and compound models.
            % structure of the methods object
            The method schema has three main attributes: associated with method parameters, method data, and precision. The \texttt{parameters} attribute holds a list of annotated control variables, which apart from the central key-value pair also encompasses a categorization keyword and, if applicable, a definition of the value's unit. The method \texttt{data} attribute contains other input variables which may require additional files, such as user-generated pseudopotentials or basis sets.

            % mention ML approach to learn precision
            % "learn" experience, what optimizer works, etc.
            
            % put figure here instead!!
            %\lstinputlisting[language=JSON]{02-methodology/example/method.json}
            
            In principle, the method schema contains all relevant information for the precision of a given choice of model and material. If one is able to formulate suitable scoring functions, the precision parameters can be turned into numeric features for a regression model. In conjunction with other factors, such as simulation time or memory usage, it would be highly desirable to predict an optimal model/method for a given material-property combination.
            % precision score

        \subsubsection{Other Entities}
        \label{subsubsec:methodology-entities-other}
            % Material structures, properties, Workflow, etc.
            In the following, we briefly outline other notable ESSE data schemas. A more detailed definition can be found in Ref.~\citenum{Bazhirov2019-ep}.
            % Material
            For materials data to be searchable, traceable, and reproducible, it is crucial to have a concise and informative way to describe materials and their properties.
            % add something about identifiers, searchability?
            The \textit{material} schema comprises descriptive properties that uniquely specify a material, such as Bravais lattice vectors and the unit cell basis. Note that the material schema is not limited to periodic systems and will also support molecular descriptors in an upcoming future release.
            % Workflow
            A workflow defines the logical composition of simulation tasks that derive from one or several simulation engines or may take other forms such as Python scripts. The workflow as we define it is also hierarchically organized in three consecutive levels (from top to bottom): workflow, subworkflow, and workflow unit. In the \textit{workflow} schema the logical composition is represented in terms of a directed acyclic graph (DAG), whereby each node is a workflow or subworkflow. A workflow may contain several subworkflows or other workflows.
            % Property
            The organization of the simulation results is managed by the \textit{properties} schema. Apart from the property data, this schema assigns a property group for easier access/findability and includes the unit of the property (if applicable).

    \subsection{Classification Rules}
    \label{subsec:methodology-classification}
        
        Since the classification and hierarchy of models involve some arbitrariness, we propose a set of classification rules to guide the categorization of hybrid models or edge cases. Although the rules listed below (Table~\ref{tab:methodology:rules}) only pertain to a part of the categorization tree, they are intended to demonstrate how specific cases can be distinguished and assigned to a category. Instead of aiming at a final set of categorization rules allowing to uniquely classify each computational model (and without discussing whether such an approach is even possible), we suggest the readers consider our approach below as a stepping stone toward a practically applicable implementation.
        
        % example of QMC
        As an example, let us consider the Quantum Monte Carlo (QMC) model - an approach to find highly accurate solutions to the quantum many-body problem and is often used to study materials and molecular systems.\cite{Foulkes2001-yx, Austin2012-zz} Due to its stochastic foundation, QMC results involve a quantifiable random error. Although stochastic sampling is an important component of the model, the aim of most QMC models is to solve for the ground state wave function (or density matrix).\cite{Foulkes2001-yx} As such the model was assigned to the \textit{ab initio} wave function model category (\texttt{abin}). This example prompts another guideline to introduce for the categorization of computational models: in case of ambiguity, one should categorize a model based on its objective (e.g. solving the Schr\"{o}dinger equation) rather than its components or derivation.
        
        % tier I-III not for actual model instances
        Another guideline concerns the relationship of categorization tiers and explicit realizations, i.e. instances of unit models.
        As outlined above, tier I to tier III serve as identifiers for groups of models.
        To guarantee a consistent usage of the unit model object, it should thus be avoided to equate a unit model instance with one of these three tiers.
        
        \begin{table}[!ht]
            \centering
            \caption{List of CateCom classification rules.}
            \label{tab:methodology:rules}
            \begin{tabular}{ccp{11cm}}\toprule
            No. & Category & Categorization Rule\\\midrule
            1 & \texttt{pb} & The model is based on physical laws.\\
            1.1 & \texttt{pb/qm} & The model depends on electronic coordinates or involves an electronic or nuclear wavefunction.\\
            1.1.1 & \texttt{pb/qm/abin} & The model is based on first-principle wavefunction approximations.\\
            1.1.2 & \texttt{pb/qm/dft} & The model is based on density functional theory.\\
            1.1.3 & \texttt{pb/qm/semp} & The model only treats valence electrons explicitly and/or involves parametrization of two-electron integrals.\\
            1.2 & \texttt{pb/at} & The model depends on atom (nuclear) coordinates only (without using wavefunctions).\\
            1.3 & \texttt{pb/mes} & The model involve a conflated representation of particles.\\
            2 & \texttt{st} & The model predicts results based on data rather than physical laws.\\
            2.1 & \texttt{st/prob} & The model involves randomness and cannot predict a result with an exact formula. Often the result is characterized by a mean and a distribution.\\
            2.2 & \texttt{st/det} & The model does not include randomness and always gives the same prediction.\\
            2.2.1 & \texttt{st/det/lin} & The model comprises a linear combination of features (or kernels).\\
            2.2.2 & \texttt{st/det/nn} & The model employs a neural network architecture.\\
            2.2.3 & \texttt{st/det/dtr} & The model is based on decision trees.\\
            \bottomrule
            \end{tabular}
        \end{table}

    \subsection{Entity Interoperation}
    \label{subsec:methodology-entity-interoperation}
\begin{figure}
            \centering
            % https://drive.google.com/file/d/1Fp-6L9LEI8kinG8umWv3nFpomNWwOnzL/view?usp=sharing
            \includegraphics[width=0.81\textwidth]{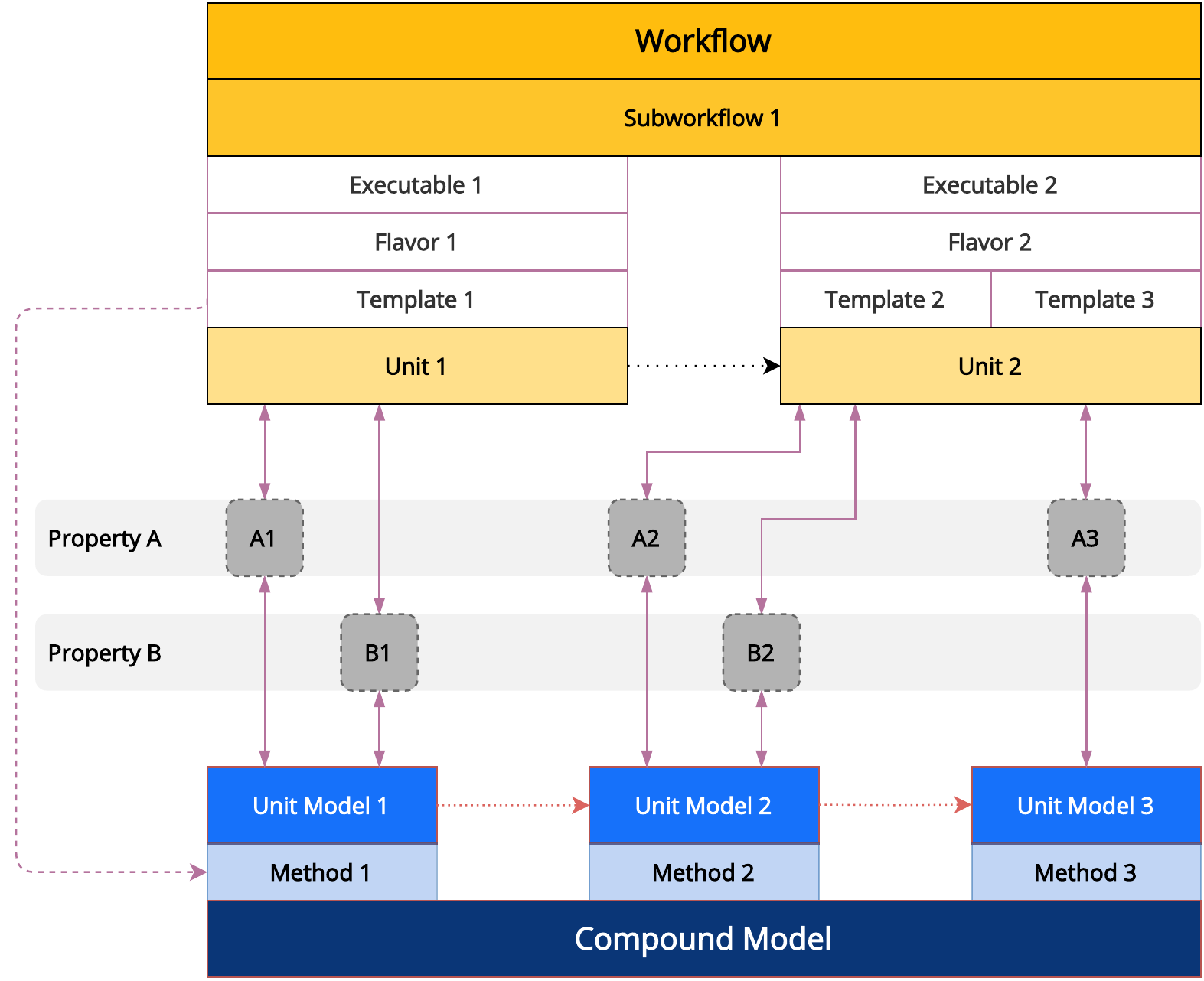}
            \caption{Interoperation of a workflow (top) and compound model (bottom). The general example involves a workflow consisting of two workflow units and a compound model with three unit models. The modular approach of unit models allows a one-to-many mapping of workflow units to unit models and tracking of intermediate properties (e.g. A1 and B1). The entities presented here follow the color scheme of Figure~\ref{fig:methodology:entities}.}
            \label{fig:examples:compound_general}
        \end{figure}

        This section describes how the entities presented in Sec.~\ref{subsec:methodology-entities} interact to facilitate the research process. In particular, we present the relation between workflow, compound model, and properties. Starting with the workflow, a hierarchical organization similar to the model entity is used. A subworkflow, which is associated with one application or software package, contains one or several workflow units. There are various types of workflow units each managing a different role, for instance, input/output operations or conditional operators (see Ref.~\citenum{Bazhirov2019-ep} for a full list of types). The workflow unit type shown in Fig.~\ref{fig:examples:compound_general} is of the \textit{execution} type, which refers to an executable of the simulation software package. An executable may require one or several input files, which are generated by templates. The \textit{flavor} entity, which is part of the workflow unit, matches templates to an executable for the purpose of obtaining a selected set of properties. In a broader sense, the flavor therefore represents a specific group of simulations, for instance, single-point calculations or geometry optimizations.
        The \textit{compound model} defines the model and simulation parameters for a given subworkflow. As outlined in Sec.~\ref{subsubsec:methodology-entities-model} each compound model is comprised of one or more unit models. Although a unit model is only associated with one workflow unit, the opposite does not apply. Allowing one workflow unit to map to several unit models makes this framework very flexible and consistent across different simulation packages. Each unit model as well as the compound model itself contains a method object which stores information related to \textit{how} a model is solved and is used to populate templates. Finally, associating properties with individual workflow units and unit models enables the user to monitor the progress of a given property across the compound model. Furthermore, since some unit models may not give rise to a certain property (cf. Unit model 3 and Property B in Fig.~\ref{fig:examples:compound_general}), the concept allows for convenient access to the last (or other criteria) occurrence of the property.

%% file: 03-examples/examples.tex
\section{Examples}
\label{sec:examples}
    
    In the present section, we elaborate the above-introduced data structures for Unit Model, Compound Model, and Method by means of common-use examples. For the sake of brevity, only a few examples are shown and we refer the reader to the ESSE repository\cite{ESSEgithub} for an extensive collection of examples. It should be noted that the reoccurring key \texttt{"\_id"} is not a component of the CateCom data structure, but pertains to the document-based database software and will thus be omitted in the discussion below. 
    
    \subsection{Unit Models}
    \label{subsec:examples-unitmodels}
    
        Kohn-Sham Density Functional Theory (KS-DFT)\cite{Hohenberg1964-ge, Kohn1965-oq} is a widely used quantum-mechanical model in material science as well as molecular science. As the solution of KS-DFT is often used as an input for a subsequent model, for instance perturbation theory\cite{Golze2019-kd} or as an alternative reference determinant\cite{Rettig2020-nu} in \textit{ab initio} wavefunction models, it is well suited to be represented as a unit model. Listing~\ref{lst:unit_model:ksdft} shows the unit model data structure for a generalized gradient approximation (GGA) KS-DFT model using the Perdew-Burke-Ernzerhof (PBE)\cite{Perdew1996-tf} exchange-correlation functional. An example of the range-separated hybrid functional HSE06\cite{Krukau2006-kv} is presented in the appendix (see Listing~\ref{lst:unit_model:ksdft:hse06}). According to the CateCom approach introduced in Sec.~\ref{subsubsec:methodology-entities-model}, the three tiers for KS-DFT are \textit{physics-based}, \textit{quantum-mechanical} and \textit{density functional theory}, respectively. Since there exist multiple realizations of DFT (for instance orbital-free DFT\cite{Wesolowski2013-ot, Witt2018-qc}), the \texttt{type} field further specifies the variation of DFT. In the data structure each tier (as well as type and subtype if applicable) is mapped to a simple object which contains a human-readable descriptor (\texttt{name}) and a machine-readable token (\texttt{slug}). The annotation fields introduced in Sec.~\ref{subsubsec:methodology-entities-model} (\texttt{augmentation}, \texttt{modifier} and \texttt{tag}) give further descriptive information about the unit model and facilitate the search for unit models. In addition, references to the literature can be given using the \texttt{reference} field (left empty in Listing~\ref{lst:unit_model:ksdft} for brevity). Each unit model includes a so-called \texttt{flowchartId} field, which is used to uniquely identify a unit model within a compound model and which serves as a reference for representing the compound model as a directed acyclic graph (DAG). 
        
\begin{lstlisting}[language=JSON, captionpos=t, caption={
            Example of a unit model data structure for Kohn-Sham Density Functional Theory. Apart from the categorization fields the data structure also holds a detailed representation of the exchange-correlation functional (here: PBE).
            }, label={lst:unit_model:ksdft}]
{
    "tier1": {
        "name": "physics-based",
        "slug": "pb"
    },
    "tier2": {
        "name": "quantum-mechanical",
        "slug": "qm"
    },
    "tier3": {
        "name": "density functional theory",
        "slug": "dft"
    },
    "type": {
        "name": "Kohn-Sham DFT",
        "slug": "ksdft"
    },
    "subtype": {
        "name": "Generalized Gradient Approximation",
        "slug": "gga"
    },
    "functional": {
        "name": "Perdew-Burke-Ernzerhof 1996",
        "slug": "PBE",
        "components": [
            {
                "name": "PBE",
                "slug": "pbe-c",
                "type": "correlation",
                "fraction": 1.0
            },
            {
                "name": "PBE",
                "slug": "pbe-x",
                "type": "exchange",
                "fraction": 1.0
            }
        ]
    },
    "tags": [
        "self-consistent"
    ],
    "method": {...},
    "reference": {...},
    "flowchartId": "u123fndff333",
    "_id": "ffffff55555"
}
\end{lstlisting}
        
        Apart from the categorization and annotation fields, the CateCom approach also supports additional fields that are exclusive to a certain unit model. For instance, the multitude of density functional approximations (DFA) warrants a separate key (termed \texttt{"functional"}), which captures the different categories of DFAs. The functional object contains identifier fields \texttt{name} and \texttt{slug}, which include the commonly used approximations and acronyms for DFAs. Many exchange-correlation functionals are comprised of several components (here referred to as \textit{unit functionals}), for instance, separate approximations for exchange and correlation as well as a fraction of exact exchange (hybrid functionals). The \texttt{functional} data structure lists these contributions under the key \texttt{components}. Each component is characterized by nominal descriptors (\texttt{name} and \texttt{slug}), the type of functional component (\textit{vide infra}) and the fraction with which the component enters the model. Besides the two unit functional types presented in Listing~\ref{lst:unit_model:ksdft} a unit functional may adopt the type of a non-separable exchange-correlation functional (e.g. GAM\cite{Yu2015-vb}), a kinetic energy functional (e.g. Thomas-Fermi\cite{Thomas1927-uc, Fermi1928-xc}), or a non-local correlation functional such as VV10\cite{Vydrov2010-dv}. 
        The \texttt{method} field in this example is left empty as it will be discussed separately in the next section.

    \subsection{Methods}
    \label{subsec:examples-methods}
    
        As mentioned in Sec.~\ref{subsubsec:methodology-entities-method}, each unit model (and compound model) comprise method data that pertains to the precision of the computational simulation. For the above example of KS-DFT, this data holds, among others, information about the employed basis (e.g. plane wave energy cutoff) and integrals (e.g. k-point grid size). The \texttt{method} data structure is organized as follows. First, a simple categorization is given by the two required fields \texttt{type} and \texttt{subtype}. The type/subtype categorization for the methods is a preliminary solution, which -if the need arises - will be replaced by a more elaborate system comparable to the unit model categorization. In the specific example of Listing~\ref{lst:method:ksdft}, the given type defines the use of plane waves in combination with pseudopotentials, while the subtype further specifies the use of ultra-soft pseudopotentials (\texttt{us}). 
        
\begin{lstlisting}[language=JSON, captionpos=t, caption={
        Data structure for the \texttt{method} entity. This example pertains to the use of plane waves and pseudopotentials for the description of the electronic structure. The data structure also includes non-default input parameters, two of which are deemed to have an effect on the overall precision.
        }, label={lst:method:ksdft}]
{
    "type": "pseudopotential",
    "subtype": "us",
    "parameters": [
        {
            "name": "ecutrho",
            "value": 1e-8,
            "categories": ["wavefunction", "precision"],
            "units": "Ry"
        },
        {
            "name": "ecutwfc",
            "value": 1e-6,
            "categories": ["wavefunction", "precision"],
            "units": "Ry"
        },
        {
            "name": "occupations",
            "value": "smearing",
            "categories": ["brillouin-zone"]
        }
    ],
    "precision": [
        {
            "name": "ecutrho",
        },
        {
            "name": "ecutwf",
        }
    ],
    "data":{
        "searchText": "",
        "pseudo": [...]
    }
}
\end{lstlisting}
        
        The method data structure also holds a list of non-default input variables in the \texttt{parameters} field. Each parameter is given in the form of an annotated key-value pair containing the name of the input variable (key) as it appears in the input file, its value, the corresponding categories (\textit{vide infra}) and, if applicable, the unit of the value. As each \textit{flavor} (cf. Sec.~\ref{subsec:methodology-entity-interoperation}) is associated with a set of default input variables, the \texttt{parameters} field only needs to store input variables which deviate from the default value. The parameters in Listing~\ref{lst:method:ksdft} stem from a plane wave DFT calculation employing the Quantum ESPRESSO software package.\cite{Giannozzi2020-gs} In particular, they define the kinetic energy cutoff for the charge density (\texttt{ecutrho}) and the wavefunction (\texttt{ecutwfc}) as well as the approach for sampling the Brillouin-zone (\texttt{occupations}). Parameters labeled with the \textit{precision} category are considered to influence the precision of the corresponding unit model and thus fulfill a special role. For example the precision score (cf. Sec.~\ref{subsubsec:methodology-entities-method}) is calculated based on these parameters. For fast access, the \texttt{precision} field collects the names of these input parameters.
        The \texttt{data} field stores additional data specific to the method. For example, in case of the plane-wave pseudopotential method, the data field contains the pseudopotentials themselves (\texttt{pseudo}). In addition, \texttt{data} contains a keyword (\texttt{searchText}) for filtering or searching the \texttt{data} attribute.
        % run ML look at methods (ML workflows)
    
    \subsection{Compound Models}
    \label{subsec:examples-compoundmodels}

        % - present two or three examples of varying complexity for Unit Models
        %   * DFT (with dispersion correction)
        %   * DFT+U
        %   * DFT + GW
        %   * TDDFT
        %   * DFTB
        %   * MP2
        %   * MP2.5 ?
        %   * CCSD(T)
        %   * QMC
        %   * QM/MM
        %   * other state-of-the art models (perhaps C-GeM or kappa-OO-MP2)
        %   * should also cover examples from other trees (semp, mes, prob, det)
        % - what about ontology components?
        %   * Instances, Classes, Attributes, Relations, Function terms, etc.
        %     See https://en.wikipedia.org/wiki/Ontology_(information_science)
        
        Following the definition of the CateCom schema, we illustrate its practical use by an example, which corresponds to established models for materials and molecules, respectively. In addition, the ESSE repository contains an extensive set of examples covering classical mechanics and machine learning models. Due to the variety of unit models also multi-level models, such as the combined quantum mechanical/molecular mechanical (QM/MM) approach\cite{Senn2009-jb}, can be realized as compound models.
        
        \subsubsection{DFT+GW Model}
            
            %GW standard for direct and inverse photoemission experiments
            % single-particle green's function, many-body perturbation theory
            % can overcome some of the most notorious deficiencies of KSDFT, such as SIE, absence of long-range polarization, and the KS band-gap problem
            % BSE, neutral excitations, similarity to TD-DFT
            Although DFT is arguably one of the most popular electronic structure models, its deficiencies inhibit an accurate simulation of some experiments, for instance, photoemission spectroscopy.\cite{Van_Schilfgaarde2006-ab} The \textit{GW} approximation provides a way to improve upon the single-particle states obtained from DFT in a perturbative fashion.\cite{Golze2019-kd} While the GW approximation allows for an accurate description of "charged excitations", i.e. electronic excitations whereby an electron is added or removed from the N-electron system, neutral excitations, which preserve the number of electrons in the system, can be described using the Bethe-Salpeter equation (BSE).\cite{Blase2020-zv} The starting point for the GW approximation are the eigenfunctions $\{\phi^{\text{KS}}\}$ and eigenvalues $\{\epsilon^{\text{KS}}\}$ of the KS-DFT mean-field Hamiltonian (Hartree-Fock or DFT). The dynamically screened Coulomb potential and the single-particle energy levels obtained from GW, in turn, are input quantities for the calculation of optical excitations using the BSE. As such, the cascade of DFT, GW, and BSE are well suited to be expressed in terms of unit models.
            
\begin{lstlisting}[language=JSON, captionpos=t, caption={
            Compound model data structure for the combination of Density Functional Theory and GW.
            }, label={lst:compound_model:ksdft:gw:bse}]
{
    "_id": "com111",
    "name": "KS-DFT + GW + BSE",
    "modelGraph": [
        {
            "name": "KS-DFT",
            "slug": "pb-qm-dft-ksdft",
            "flowchartId": "abcd1234",
            "next": "efgh5678",
            "head": true,
            "workflowUnitId": "d41a7b69cd9696"
        },
        {
            "name": "GW",
            "slug": "pb-qm-abin-gw",
            "flowchartId": "efgh5678",
            "next": "ijkl91011",
            "head": false,
            "workflowUnitId": "f790d4fbaac5e1"
        }
        ,
        {
            "name": "BSE",
            "slug": "pb-qm-abin-bse",
            "flowchartId": "ijkl91011",
            "next": null,
            "head": false,
            "workflowUnitId": "f790d4fbaac5e1"
        }
    ],
    "method": {...},
    "tags": [
        "excited-states",
        "Green's function"
    ]
}
\end{lstlisting}
            Turning to the compound model data structure (Listing~\ref{lst:compound_model:ksdft:gw:bse}), a slightly different object composition can be seen. Since most of the information pertaining to the overall model is stored within the unit model data structures, the compound model does not need to repeat the information in its own data structure. Consequently, the compound model holds the arrangement of unit models (\texttt{modelGraph}) and a global methods object (\texttt{method}).
            The \texttt{modelGraph} field contains a list of nodes, each representing a unit model. Each node contains the necessary fields to construct a directed acyclic graph, i.e. a unique identifier (\texttt{flowchartId}), a pointer to the next object (\texttt{next}) and an boolean indicator of the first node (\texttt{head}). In addition, a human-readable (\texttt{name}) and a machine-readable (\texttt{slug}) label are included. Finally, each unit model node also maps to a workflow unit by means of the \texttt{workflowUnitId} key, which does not necessarily have to be unique to each node. For instance, in the above example, three unit model nodes are mapped to two workflow units. A possible scenario for this is calculating the KS-DFT solution from one software package (e.g. Quantum ESPRESSO) and subsequently applying GW and BSE using another (e.g. BerkeleyGW\cite{Deslippe2012-dy}). Just like the unit model, the compound model data structure also includes a \texttt{method} key, which refers to a \textit{global} method configuration.
            
            \begin{figure}[ht]
                \centering
                % https://drive.google.com/file/d/1Pt0gfVJDQUErXC6siwrkbhIN19vj57To/view?usp=sharing
                \includegraphics[width=0.95\textwidth]{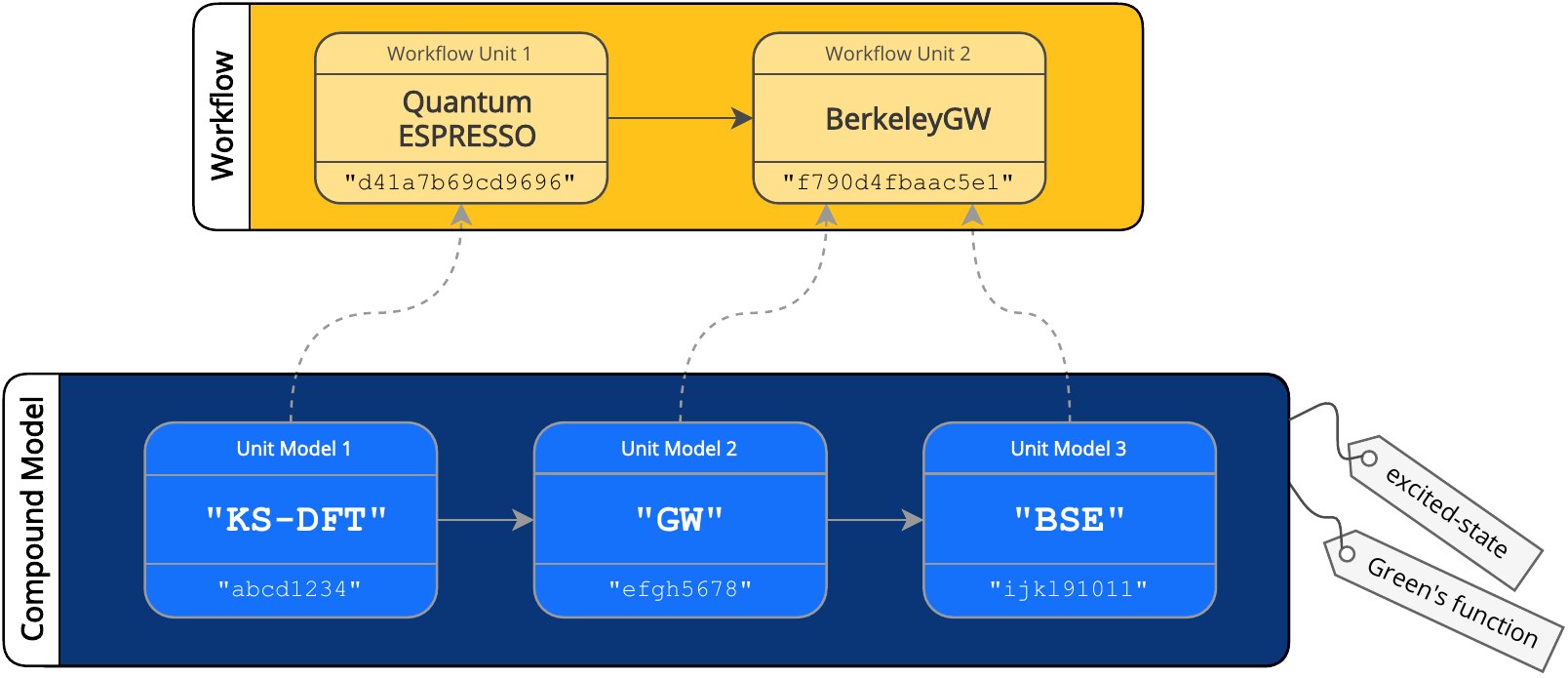}
                \caption{
                    Relation of compound model and workflow for the many-body theory example of GW and Bethe-Salpeter Equation (BSE) based on a reference from Kohn-Sham density functional theory (KS-DFT). In this example, the calculation of KS-DFT reference and application of many-body theories is performed in two steps and thus two independent workflow units.
                }
                \label{fig:examples:dft_gw_bse}
            \end{figure}
            
        %\subsubsection{Machine Learning}

%% file: 04-discussion/discussion.tex
\section{Discussion}
\label{sec:discussion}

    The CateCom approach laid out in previous sections introduces a model data structure combined with a systematic categorization of computational models. The model data structure is designed to be composed of one or more reusable components (unit models), each of which is assigned a model category. The model and other Entities representing the computational Workflow, Properties, and Materials altogether form the research-work-related data and metadata. Our approach is not meant as a final "polished" solution, but rather as a proof-of-concept but practically deployable implementation. Admittedly, it is largely limited to physics-based models in the current implementation and is, perhaps, heavily biased toward atomistic and nanoscale simulations. The uniqueness of categorization - whether to enforce it and how - is another topic that requires further clarification. Without uniqueness, the chosen categorization scheme can be seen as one linear realization of a non-linear "model graph". The rest of this section discusses several important aspects of CateCom.
     
    \subsection{The Material-Model-Property Categorization}
    
        % findability
        \subsubsection{Material, Model, and Property relationship}
        For any practical applications, the fidelity of the modeling approaches can usually only be established within a certain class of materials and their associated properties. For example, it is well known that conventional Density Functional Theory significantly under-estimates the electronic band gap values in semiconductors, while providing adequate predictions for other properties such as lattice constants and/or vibrational spectra. Therefore a certain fidelity metric predicting the quality of a particular model would only stand with respect to certain material and property types.
        
        % findability
        \subsubsection{Material and Property Categorization}
        To facilitate data-driven science, a coupled approach is needed where materials (and chemicals) have to be categorized as well as their derived properties. This way one can construct the associative relationships that can assist in identifying the most successful combinations of Materials, Workflow/Model, and Properties. Following the example in the previous section, such an approach should be able to identify, for example, that for III-V semiconductors a Model/Workflow containing both Density Functional Theory and GW Approximation provides higher fidelity than Density Functional Theory alone. Although the exact nature of such categorization is a topic of a separate discussion, and the categorization related to each of the entities can be interdependent, we still believe that starting with the model categorization provides a viable and practically useful first step.
    
    \subsection{FAIR Principles}
    \label{subsec:discussion-fair}
    % Explain the use of CateCom for building the FAIR ecosystem
    % findable: slugs, tags, flowchartId (metadata), structure
    % accessible: ESSE github, open source 
    % interoperable: interplay of entities, different software packages
    % reusable: unit models (even with different implementations)
    % some general words about the FAIR ecosystem
    There have been many efforts to collect and systematically organize research data to build large publicly available datasets.\cite{Curtarolo2012-ik, Jain2013-va, Saal2013-ko, Calderon2015-vr, Kirklin2015-qz, Draxl2018-mm, Choudhary2020-uc, Huber2020-mv} Simultaneously, a new paradigm for scientific discovery, data-driven science, emerged, which aims to detect patterns or anomalies in these types of datasets.\cite{Hey2009-rw, Draxl2018-mm, Draxl2019-yj} In a cooperative effort including representatives from academia, industry, funding agencies, and scholarly publishers, the FAIR guidelines\cite{Wilkinson2016-rq, Wilkinson2018-ry} (findable, accessible, interoperable, and reusable) were developed in order to enhance data reusability. The following subsections demonstrate how the CateCom approach ties in with the FAIR principles.

        % findability
        \subsubsection{Findability}
        According to the FAIR guiding principles\cite{Wilkinson2016-rq}, findability involves the use of globally unique and persistent identifiers. As described in Sec.~\ref{subsec:examples-unitmodels}, each unit model in the CateCom scheme encompasses such a unique identifier in the form of the \texttt{flowchartID}. Additionally, unit models and method parameters are enriched with \texttt{tags} metadata facilitating a search or filtering for certain properties. In this way, associations between unit models can be made, which are not represented in the CateCom tree. For instance, the \texttt{scaling-power-3} tag may be used to filter all unit models which formally exhibit a cubic scaling even if they are located in different categorization branches.
    
        % accessibility
        \subsubsection{Accessibility}
        One aspect of the FAIR guidelines pertains to the accessibility of the data and the protocols applied in that process. The CateCom approach implements database schema corresponding to the Unit Model, Compound Model, or workflow using the JSON schema vocabulary. As such, the resulting Entity objects are represented in the widely adopted JSON format. The model Entities are stored in a document-based database and are thus easily retrievable using the unique database identifier or through a database search. The CateCom schemas themselves are publicly available since they are part of the open-source ESSE repository\cite{ESSEgithub}. In addition, the JSON format is widely used on the world wide web in many web applications. The latter implementation provides packages for two of the currently most widely used languages - Python and JavaScript - with an intent to allow for easy adoption in software development efforts providing user interface components that in turn aid Accessibility.
    
        % interoperable
        \subsubsection{Interoperability}
        Interoperability encompasses the integration with other data and cross-functional cooperation with applications. The CateCom Unit Models do not possess dependencies to specific software implementations of the models, such that the models can, in principle, be associated with any software package (given that the model is implemented therein). Furthermore, storing the entities as JSON objects has the advantage that there are plenty of resources available which directly accept this format or are able to convert it to a different format (see also Sec~\ref{subsec:discussion-community}). The software implementation mentioned in the previous section provides additional opportunities for building interoperable systems.
        
        % reusable
        \subsubsection{Reusability}
        The CateCom scheme also addresses the reusability of data. In particular, the partitioning of models into unit models serves the purpose of reusing components that make up a model. A good example for this are many-body perturbation theory models which generally require the solution to an unperturbed Hamiltonian $\hat{\mathcal{H}}_{0}$. Consequently, the Unit Model corresponding to the unperturbed Hamiltonian can be combined with different perturbation theory models or, in the case of Hartree-Fock theory, with post-HF wavefunction models such as configuration interaction (CI). Furthermore, the storage of the method data plays a crucial part in recording the provenance of the final property data.
        %As suggested in Sec.~\ref{subsubsec:methodology-entities-method} this information 
    
    \subsection{Predictive AI/ML}
    \label{subsec:predictive-ai-ml}
    
        \subsubsection{Avoidance of Duplicates}
        With the ability to quantify and store the metadata about the digital approaches comes the ability to avoid repetition. In case a particular workflow/model/property combination, for example - pseudopotential Density Functional Theory with a certain wavefunction and charge density cutoffs - has been applied to a specific material, our proposed data management model will be able to provide a way to generate a unique fingerprint. Based on such a unique fingerprint any further duplicate attempts can be avoided leading to improved efficiency of the research work.
        
        \subsubsection{The AI "Chemist-in-the-cloud"}
        The ultimate goal of the categorization described here is to enable the creation of an AI-powered digital computational chemist/materials scientist (“brain”) able to suggest the best model/method combinations for characterizing materials. The complexity of materials science and chemistry is to a large degree defined by the diversity of the problem sets and the parametric conditions associated with them. Although computational techniques have been around for over half a century, the ability to apply them successfully with high fidelity still has a significant "art" component requiring very specialized knowledge limited to a select group of scientists only. And even this select group in practical applications often relies on their "intuition" derived through years of experience in dealing with specific problems rather than purely deterministic. With the help of the categorization scheme proposed and provided sufficient training data based on expert decisions such intuition can be instead represented as data-driven AI/ML approaches.

    \subsection{Community, Ecosystem, and Future Outlook}
    \label{subsec:discussion-community}
    
        \subsubsection{The Global Digital Ecosystem for Materials R\&D}
        The categorization framework and associated ontologies both represent fundamental critical steps in the implementation of a global digital ecosystem for materials R\&D. Having data standards is an important fundamental step in the design and implementation of the data- and software infrastructure for such an ecosystem. Object-oriented design for the entities and data structures naturally enables modularity when building the software components of such ecosystem, and greatly streamlines its implementation and long-term maintenance. A version of the present categorization framework have been previously deployed as part of an online software platform \cite{exabyte} with applications demonstrated for multiple use cases, including metallic alloys \cite{2016-exabyte-aps-abstract}, electronic properties of semiconductors \cite{2018-exabyte-accessible-CMD, 2018-exabyte-binary-compounds}, vibrational properties of materials \cite{2018-exabyte-phonon}, adsorption and catalysis in zeolites\cite{2019-chehaibou-bazhirov-jctc-zeolites}, adhesive strength of composite materials\cite{2019-koyanagi-bazhirov-adv-composite-materials}, and beyond.
        Materials R\&D spans a complex and multi-dimensional landscape, and requires an extremely large variety of characterization data at multiple time- and length scales. Once obtained, the data must be stored and managed in an efficient way. As more and more of materials research is performed in a way that involves digital handling of data, ontologies and categorization becomes important. This will facilitate the availability of ever increasing amounts of materials data on the web with contributions from the global community.
    
        \subsubsection{Community Contributions}
        % Explain how to contribute
        % github ESSE
        % needs experts mapping out the model landscape
        % refine and expand unit model tree
        % build library of compound models
        % forking workflow
        % In addition to the in-depth understanding of a model, categorization also calls for
        Simulation scientists are able to resort to a myriad of statistical and physics-based models, whereby the number of models is constantly growing. This circumstance makes a systematic mapping of the "model landscape" rather difficult for a small team since profound knowledge of a model is required in order to systematically arrange its properties and variants. Of course, expert knowledge is also indispensable for identifying reusable components of these models and examining edge cases that may fit more than one category. Thus, a more effective approach is to involve the community of experts directly in the maintenance and expansion of the categorization scheme. To this end, we propose to follow the collaborative strategy typical for code development platforms such as GitHub. These platforms also allow interested contributors to discuss new features (e.g. a new category) and raise issues about existing ones. In practice, a contributor first obtains their own server-side clone ('fork') of the original repository (e.g. ESSE\cite{ESSEgithub, Bazhirov2019-ep}). The implementation of new features, such as a new unit model, is then carried out in a feature branch located in the cloned repository. Once the new feature is ready, the contributor then issues a request for the integration of the new feature ('pull request') to the maintainer of the original repository. At this stage details of the new feature can be discussed and modified until the maintainer accepts the incoming changes.
        
        \subsubsection{Interfacing with other approaches}
        Of course, the task of developing a global digital ecosystem for materials research and development cannot be accomplished without involving a global community and interfacing with other efforts. Despite recent comprehensive approaches\cite{Andersen2021-eg, Li2020-rf}, it is still common for the materials science community to develop standards which are tailored to a specific sub-branch of research. Such "artisanal"\cite{Pizzi2016-dc} approach has led to several competing standards with a relatively small impact on the field. Our goal is to provide a common denominator allowing the key contributors to realize their ambitions while at the same time facilitating the level of quality required for practical real-world applications.
        
        % ontologies general -> semantic enrichment, translation, mapping
        CateCom has a natural connection to the principle of an ontology and the similarities facilitate interoperation of CateCom with ontologies defining model classes, for the ontology-based data access (OBDA).\cite{Calvanese2007-wv, Araujo2017-cc, Konys2017-be} Since ontology vocabulary is based on different formats (RDF, OWL, etc.), OBDA usually requires an access interface translating queries and responses.\cite{Araujo2017-cc, Konys2017-be} Ontologies may also be used for semantic annotation, i.e. metadata enrichment.
        
        \subsubsection{Example Interfaces with other approaches}
        
        The Materials Design Ontology (MDO) defines a \textit{Computational Method} class which as of this writing is limited to density functional theory (DFT) and Hartree-Fock (HF) theory. Nonetheless, the definition of these models shares commonalities with the CateCom scheme. For instance, the DFT class of MDO has a \textit{Exchange Correlation Energy Functional} property implementing the most common groups of density functional approximations which are also supported by the KS-DFT unit model in CateCom. Part of the CateCom unit model (in JSON format) can thus be mapped to RDF format in order to be used with MDO. A similar mapping to the RDF format using a SPARQL-Generate script\cite{Lefrancois2017-sr} has been described in Ref.~\citenum{Li2020-rf}. Furthermore, a model class is also defined as part of the Elementary Multiperspective Material Ontology (EMMO).\cite{EMMO} Although specific models (e.g. DFT) are not explicitly represented, its subclasses are similarly organized as tier I and tier II of the CateCom scheme, for instance \textit{DataBasedModel} and \textit{PhysicsBasedModel} exactly correspond to the \texttt{st} and \texttt{pb} categories (cf. Figure~\ref{fig:model_tiers}). Interoperation of EMMO and CateCom could, for instance, be achieved by annotation or translation as described above.
        
        % Databases, mapping, JSON-to-JSON transformation (Jolt), OPTIMADE (API)
        In practical terms, any interfacing most likely involves a conversion between two database schemas. The Novel Materials Discovery (NOMAD\cite{Draxl2018-mm}) metainfo schema defines the structure of material-science-related data. The schema contains a very extensive list of entities and properties, such that mapping is not limited to CateCom unit models but can in principle extend to other ESSE\cite{ESSEgithub, Bazhirov2019-ep}, Entities (e.g. Method, Property). Apart from a one-to-one mapping, one could also populate the CateCom data structures based on a descriptive string. For instance, JARVIS-DFT database\cite{Choudhary2020-uc} contains a \textit{functional} property, which contains enough information in order to be converted to a \texttt{ksdft} CateCom model object. Such object generation might not be fully complete, neglecting the method object entirely. Other approaches such as OPTIMADE\cite{Andersen2021-eg} provides a universal application programming interface (API) to access material data across several databases. The OPTIMADE specification always includes a structure attribute, whereas properties other than structural or chemical information are provider-specific. As a consequence, model-related metadata (and therefore the mapping to CateCom) may or may not be available.
        
        \subsubsection{Future Outlook}
        
        In our vision, CateCom presents a fundamental building block in facilitating mainstream data-driven research in materials science and chemicals. Our goal is to engage a large community of people possessing specialized knowledge about materials and chemicals in digital work resulting in the creation of novel AI/ML techniques. We see that community effort is critical in obtaining the "critical mass" of data and creating network effects allowing to sustain the effort for the long term. To understand the future outlook, we draw analogies with the Computer-Aided Design and Electronic Design Automation industries. In both, as the transition from exploratory science-centric to practically applied engineering-focused research work was progressing, the number of data representation standards was consolidated to 3-5. These consolidated standards emerged behind the software development efforts amassing the largest user community - such as AutoDesk, Synopsys, Dassault, etc. We expect a similar progression of events to happen for data-driven digital materials R\&D in the near future.
        
        Apart from the general goal, there are also more technical questions that could be addressed in future work. More specifically, the current approach allows the combination of unit models to compound models without any restrictions. In order to prevent improper combinations, a strategy for interfacing unit models is needed. A potential solution would be to track input and output quantities of each model, such that combinations can be evaluated in terms of the intersection of input and output quantities.
        Another point concerns the representation of time-dependent simulations. Future work should thus examine whether unit models may include a "time-propagation" operator or whether a compound model analog (e.g. "dynamical compound model") would be a viable option.
        
        %Further work on the categorization effort is needed, which we hope to achieve together with the community of computational scientists.
        Regarding the categorization, it would be desirable to extend an existing ontology or create a new ontology for the multitude of computational models spanning both physics-based and data-driven models. Such an ontology would be helpful for building a knowledge graph of materials science research containing the semantic relationships between material, model, and property entities. 

%% file: 05-conclusion/conclusion.tex
\section{Conclusion}
\label{sec:conclusion}
% - short summary of CateCom
%   * open source
%   * object-oriented, modularized
%   * 
% - encourage community contribution
% - outlook (what can still be improved?, what will it be used for?)
    We introduced an approach for the categorization of computational models in conjunction with database schemas representing the Models and Methods. The proposed data-centric categorization scheme follows an object-oriented design concept, whereby a given model is expressed in terms of reusable, indivisible components (Unit Models). This modular unit model approach allows for a consistent description of model properties across software packages and is able to describe multi-level models, such as QM/MM.
    % examples
    The data structures derived from the proposed schemas have been elucidated based on the examples, such as Density Functional Theory (DFT), GW Approximation, and similar. It has been demonstrated how the CateCom scheme complies with the FAIR guiding principles. In particular, possible mechanisms for the interoperation of CateCom with other approaches of the digital materials science ecosystem have been presented.
    % uniqueness / completeness
    In order to manage limit cases and guide new additions of categories, a set of categorization rules has been presented. 
    
    In its current state of development, CateCom represents a proof-of-concept with an emphasis on physics-based models. With the aim of leveraging expert knowledge, we discussed a community-driven approach for the extension of the CateCom scheme. Just as many other categorization efforts the CateCom scheme does not claim uniqueness with regards to the chosen categories.
    % Benefits of classification
    The organization of model data as presented herein allows for several convenient benefits such as transferability of a given model from one problem to another. The model categorization also allows for the generation of unique fingerprints, which facilitate the research process by avoiding duplicates. We share the current implementation of the categorization as part of an open-source online codebase\cite{ESSEgithub} and demonstrate some of the applications of the underlying data infrastructure in the online platform\cite{Bazhirov2019-ep}.

    The ideas expressed in the present manuscript build upon the Materials Genome Initiative\cite{materials-genome-initiative-website}, and are designed to facilitate collaboration between materials scientists, chemists, computer/data scientists to create, deploy and analyze a set of curated methodologies to rapidly study materials at multiple time- and length scales. In our view, the present data convention is aimed to facilitate the next generation of computer-aided design tools and enables advanced R\&D capabilities that facilitate the development of new kinds of products in critical industries including semiconductor, photovoltaics, energy storage, oil \& gas, specialty chemicals, aerospace and automotive and others and has the potential to transform the materials sector at large.

%% file: 99-supporting-info/supporting_information.tex
\begin{lstlisting}[language=JSON, captionpos=t, caption={
    Unit model example for Kohn-Sham density functional theory using the HSE06 exchange-correlation functional.},
    label={lst:unit_model:ksdft:hse06}]
{
    "tier1": {
        "name": "physics-based",
        "slug": "pb"
    },
    "tier2": {
        "name": "quantum-mechanical",
        "slug": "qm"
    },
    "tier3": {
        "name": "density functional theory",
        "slug": "dft"
    },
    "type": {
        "name": "Kohn-Sham DFT",
        "slug": "ksdft"
    },
    "subtype": {
        "name": "Hybrid functional",
        "slug": "hybrid"
    },
    "functional": {
        "name": "HSE06",
        "slug": "hse06",
        "components": [
            {
                "name": "PBE",
                "slug": "pbe-x",
                "type": "exchange",
                "fraction": 0.75,
                "range": "short-range"
            },
            {
                "name": "Exact exchange",
                "slug": "hf",
                "type": "exchange",
                "fraction": 0.25,
                "range": "short-range"
            },
            {
                "name": "PBE",
                "slug": "pbe-x",
                "type": "exchange",
                "fraction": 1.0,
                "range": "long-range"
            },
            {
                "name": "PBE",
                "slug": "pbe-c",
                "type": "correlation",
                "fraction": 1.0
            }
        ]
    },
    "tags": [
        "user-adjustable",
        "range-separated"
    ],
    "method": { ... },
    "reference": { ... },
    "flowchartId": "u123fndff444",
    "_id": "ffffff666666"
}
\end{lstlisting}

\begin{lstlisting}[language=JSON, captionpos=t, caption={
    Unit model example regarding coupled cluster theory including singly and doubly excited determinants (CCSD).
    },label={lst:unit_model:abin:cc}]
{
    "tier1": {
        "name": "physics-based",
        "slug": "pb"
    },
    "tier2": {
        "name": "quantum-mechanical",
        "slug": "qm"
    },
    "tier3": {
        "name": "ab-initio",
        "slug": "abin"
    },
    "type": {
        "name": "coupled cluster",
        "slug": "cc"
    },
    "subtype": {
        "name": "CC Singles Doubles",
        "slug": "ccsd"
    },
    "tags": [
        "single-reference",
        "ground-state"
    ],
    "method": { ... },
    "reference": { ... },
    "flowchartId": "u123fndff444",
    "_id": "ffffff666666"
}
\end{lstlisting}
    
\begin{lstlisting}[language=JSON, captionpos=t, caption={
    Unit model example for the ordinary least-squares regression model from the statistical-based model branch.
    },label={lst:unit_model:lin:ols}]
{
    "tier1": {
        "name": "statistical",
        "slug": "st"
    },
    "tier2": {
        "name": "deterministic",
        "slug": "det"
    },
    "tier3": {
        "name": "linear models",
        "slug": "lin"
    },
    "type": {
        "name": "ordinary least-squares",
        "slug": "ols"
    },
    "tags": [
        "supervised",
        "regression"
    ],
    "method": { ... },
    "reference": { ... },
    "flowchartId": "u123fndff444",
    "_id": "ffffff666666"
}
\end{lstlisting}